# Reliable Real-time Seismic Signal/Noise Discrimination with Machine Learning

**Men-Andrin Meier[1], Zachary E. Ross[1], Anshul Ramachandran[2], Ashwin Balakrishna[2], Suraj Nair[2], Peter Kundzicz[2], Zefeng Li[1], Jennifer Andrews[1], Egill Hauksson[1], Yisong Yue[2]**

[1] Seismological Laboratory, California Institute of Technology, Pasadena, California, USA.

[2] Computing and Mathematical Sciences, California Institute of Technology, Pasadena, California, USA.

Corresponding author: Men-Andrin Meier (mmeier@caltech.edu)

## Key Points:

- Rapid and reliable signal/noise discrimination is one of the most important challenges for current Earthquake Early Warning (EEW) systems
- We train 5 machine learning classifiers of variable complexity using a large data base of real earthquake and impulsive noise signals
- Deep neural networks outperform shallower architectures and reach classification accuracies that may facilitate nearly error-free EEW

## Abstract

In Earthquake Early Warning (EEW), every sufficiently impulsive signal is potentially the first evidence for an unfolding large earthquake. More often than not, however, impulsive signals are mere nuisance signals. One of the most fundamental - and difficult - tasks in EEW is to rapidly and reliably discriminate real local earthquake signals from all other signals. This discrimination is necessarily based on very little information, typically a few seconds worth of seismic waveforms from a small number of stations. As a result, current EEW systems struggle to avoid discrimination errors, and suffer from false and missed alerts. In this study we show how modern machine learning classifiers can strongly improve real-time signal/noise discrimination. We develop and compare a series of non-linear classifiers with variable architecture depths, including fully connected, convolutional (CNN) and recurrent neural networks, and a model that combines a generative adversarial network with a random forest (GAN+RF). We train all classifiers on the same data set, which includes 374k local earthquake records (M3.0-9.1) and 946k impulsive noise signals. We find that all classifiers outperform existing simple linear classifiers, and that complex models trained directly on the raw signals yield the greatest degree of improvement. Using 3s long waveform snippets, the CNN and the GAN+RF classifiers both reach 99.5% precision and 99.3% recall on an independent validation data set. Most misclassifications stem from impulsive teleseismic records, and from incorrectly labeled records in the data set. Our results suggest that machine learning classifiers can strongly improve the reliability and speed of EEW alerts.

## Plain Language Summary

Seismic stations record not just earthquake signals, but also a wide variety of nuisance signals. Some of these nuisance signals are impulsive and can initially look very similar to real earthquake signals. This is a problem for Earthquake Early Warning (EEW) algorithms, which sometimes misinterpret such signals as being real earthquake signals, and which may then send out false alerts. For each registered impulsive signal, EEW systems need to decide (or, classify) in real-time whether or not the signal stems from an actual earthquake. State-of-the-art machine learning (ML) classifiers have been shown to strongly out-perform more standard linear classifiers in a wide range of classification problems. Here analyze the performance of a variety of different ML classifiers to identify which type of classifiers leads to the most reliable signal/noise discrimination in an EEW context. We find that we can successfully train complex deep learning classifiers that can discriminate between nuisance and earthquake signals very reliably (accuracy of 99.5%). Less complex ML classifiers also outperform a linear classifier, but with significantly higher error rates. The deep ML classifiers may allow EEW systems to almost entirely avoid false and missed signal detections.

# 1 Introduction

Only a few seconds typically separate the onset of an earthquake from the time at which strong shaking begins at the earth's surface. In more rare cases it can take a few tens of seconds for strong motion to arrive at a target site. This leaves very little time for Earthquake Early Warning (EEW) systems to detect events and provide useful ground motion alerts. As a consequence, useful alerts are necessarily based on very little information.

One particularly difficult challenge in providing EEW alerts is reliably discriminating between real local earthquake signals and any other kind of signal, based on only a few seconds of seismic waveform data. If reliable discriminations could be made based on only the first station that has recorded the beginning of an earthquake, most sites could be alerted in time. However, there is an inherent trade-off between alerting reliability and speed. Single station alerts are risky (e.g. Xu et al., 2016, Böse et al., 2009b) because seismometers also record a very wide range of non-earthquake signals, some of which can resemble those of earthquakes, at least in the first few seconds. Alerts based on only one station can therefore have high rates of false alerts. Many EEW systems wait until a minimum number of stations report potential earthquake signals. This strongly decreases false alert rates but at the cost of slower alerting speed and a corresponding blind zone where alerts arrive too late. This is particularly problematic in places like southern California where earthquakes occur in close proximity to populated areas.

The ShakeAlert EEW system for the US West Coast (Given et al., 2014, Kohler et al., 2018) requires detections on at least 4 sites to issue an alert. Despite this requirement, false event declarations occur several times per year (Cochran et al., 2018). In part, these false declarations are caused by misinterpretations of impulsive nuisance signals, e.g. from anthropogenic noise sources, instrument malfunctions or teleseismic signals. The OnSite algorithm for instance (Böse et al., 2009a) that until recently was part of the ShakeAlert system uses an STA/LTA filter to detect impulsive signals, and then applies a linear two-feature classifier based on the proposition of Böse et al., 2009b. It computes peak amplitudes and predominant period estimates and maps these into a "q-value", a degree of belief that the signal is a local earthquake signal. This classifier successfully discards most false detections, but on average, 184 times per day a non-earthquake signal is mistakenly assigned a non-zero q-value. If such misinterpretations happen to occur simultaneously at multiple stations in a short time window, false event declarations can occur.

The recent advances in machine learning, from both outside and within the seismological community, have strong potential to improve real-time seismic signal classification. Machine learning algorithms, in particular deep neural networks, have recently been highly successful in a wide range of tasks, including visual understanding (e.g. Deng et al., 2009, Donahue et al., 2015), natural language processing (e.g. Peters et al., 2018), and robotic navigation (e.g. Zhu et al., 2017) and control (e.g. Levine et al., 2017). In seismology machine learning has been in use for a long time (e.g. Wang et al., 1995, Böse et al., 2008) and have recently become very useful for a wide range of tasks, including signal detection (Ross et al., 2018b, Chen 2018, Rong et al., 2018, Yoon et al., 2015) and hypocenter location (Perol et al., 2018, Panakkat and Adeli, 2009), body wave arrival time picking (Zhu and Beroza, 2018) and first motion polarity assignment (Ross et al., 2018a), the prediction of failure times in laboratory experiments (Rouet-Leduc et al., 2017), seismogram encoding (Valentine and Trampert, 2012) ground motion amplitude prediction (Trugman and Shearer, 2018) and seismicity forecasting (DeVries et al., 2018).

For the signal/noise discrimination problem that EEW algorithms face, supervised machine learning algorithms similarly promise significant improvements (e.g. Maggi et al. 2017; Sermanet et al. 2014; Hammer et al., 2012). With sufficient training data, non-linear classifiers such as those based on convolutional neural networks (CNN) can usually outperform simple linear classification schemes (e.g. Mousavi et al. 2016, Kong et al., 2016). Ross et al. 2018b have demonstrated how deep learning approaches can distill the generalized characteristics of seismic body waves from a large set of labeled example seismograms. A CNN can be trained to discriminate between P- and S- body waves and ambient background noise signals with high reliability. Li et al., 2018 have combined a generative adversarial network (GAN) with a random forest classifier to discriminate between direct seismic P-phases and impulsive noise sources. Both approaches take seismic waveforms as direct input data and output the probability that a given signal belongs to any of a number of predefined signal classes. Such approaches may allow EEW algorithms to trigger only on waveforms that have the general characteristics of direct seismic body waves, rather than triggering on any impulsive signal.

In this study we develop machine learning classifiers that are optimized for reliable signal/noise discrimination in an EEW context. We have compiled a 3-component waveform database that contains 374k local earthquake records (M3.4 - 9.1, hypocentral distances up to 1,000km), 946k impulsive nuisance signals and 7.5k impulsive teleseismic earthquake records. We design and compare several different types of machine learning classifiers, ranging from simpler fully connected neural networks to convolutional neural networks. We train and evaluate all classifiers with the same input data to facilitate a direct comparison between the different approaches.

With sufficient training data the deeper architectures should in principle outperform the simpler models. The amount of currently available seismic data, however, is at the lower end of what is necessary to train deep networks effectively. Furthermore, some of the most notoriously difficult signals (e.g. signals from deep teleseismic events, or from large magnitude events) are relatively rare and we only have small numbers of training recordings, posing challenges to deep learning approaches. The classification models compared in this study will shed light on what architectures lead to a maximally powerful classifier with the limited data that is currently available.

In section 2 we describe the data set used in this study. Section 3 introduces the different classifiers. In section 4 we analyze and compare their performance. In section 5 we discuss the potential of these ML-based classifiers to significantly improve the robustness of EEW algorithms.

## 2 Data Set

The data set includes two main seismic signal classes: local earthquake records ("quake") and impulsive nuisance signals ("noise") that were not caused by local earthquakes. We apply the same waveform processing to all waveforms. We use a 2nd order causal Butterworth high-pass filter with a corner frequency of 0.075Hz and extract a set of 25 waveform features on increasingly long waveform snippets. The features are seismologically motivated in that they are quantities we think might be diagnostic of whether a signal is caused by a real earthquake. Some features were computed on the raw, i.e. unfiltered, waveforms, since some nuisance signals, such as boxcar-like signals from instrument malfunctions, are more characteristic in this form. A description of all features is given in Table 1. The snippets start at the onset of the impulsive waveform signal, here defined by the time of an automated pick, and end 1s, 2s and 3s after that. For the deep networks, which use the waveforms directly rather than waveform features, we cut the waveforms at 1s before until 3s after the impulsive onsets. We do not impose a minimum signal/noise threshold.

| Name | Abbreviation | Input data | Description |
|---|---|---|---|
| Peak absolute acceleration | pa | 3C vector sum of HP acc. | Peak absolute amplitude since P-onset |
| Peak absolute velocity | pv | 3C vector sum of HP vel. | Peak absolute amplitude since P-onset |
| Peak absolute displacement | pd | 3C vector sum of HP dsp. | Peak absolute amplitude since P-onset |
| Filter bank peak absolute amplitudes | fbamps1 - fbamps9 | HP vertical vel. | Peak absolute amplitudes in 9 octave-wide passbands between 09375Hz - 48Hz |
| Vertical/horizontal ratio | zhr | HP vel. | Peak absolute amplitude on the vert. comp. divided by peak abs. amp. of vector sum of 2 horizontal comp. |
| Zero crossing rate | zcr & zcrR | HP vel. and raw waveform | No. of zero crossings divided by signal duration |
| Skewness | skew & skewR | HP vel. and raw waveform | Skewness |
| Kurtosis | kurt & kurtR | HP vel. and raw waveform | Kurtosis |
| Squared skewness and kurtosis | k2 | HP vel. | Sum of squared skewness and squared kurtosis |
| Cumulative absolute velocity | cav & cavR | HP vel. and raw waveform | Integrated absolute velocity |
| Quarter comparison | qtr & qtrR | HP vel. and raw waveform | Median abs. amp. in last quarter of waveform snippet divided by median abs. amp. in first quarter |
| Maximum step between consecutive samples | maxstepR | Raw waveform | Max. jump between any two neighbouring time series samples |
| Pre-signal noise level | presig | HP acc. | Sample standard dev. of amplitude distribution at 1.0 - 0.5s before P-onset |
| Predominant frequency | tauC | HP dsp. and vel. | Square root of ratio of integrated squared dsp. and integrated squared vel. |
| Variance ratio | rvar | HP acc. | Ratio of sample variances in the time intervals [0 : 0.2]s after signal onset, and [0.2 : 0.4]s after signal onset |
| Extreme value | f38 | HP acc. | Max. abs. deviation from mean, divided by variance |

**Table 1** *Features used for all classifiers, except for the CNN and GAN+RF. All used digital filters were 2nd order causal Butterworth filters. Some features were computed both on processed and on unfiltered waveforms ("raw", only running mean removed and scalar gain correction, and denoted with a capital "R" in the abbreviated feature name).*

### 2.1 Local earthquake records ("quake")

The local earthquake waveform data set is an updated version of the data set of Meier (2017). It includes broadband and strong motion records from the Southern California Seismic Network (SCSN), and records from the Japanese K-NET and KiK-net strong motion networks (surface stations only). The SCSN data contains 107k records from the time period of January 1990 to November 2016, with magnitudes in the range Mw 4-7.3 and hypocentral distances from 0 - 360km. The Japanese data contains 266k records from May 1996 to October 2017, magnitudes 3.4-9.1 and hypocentral distances from 0 - 1,000km. From both regions we have included all available data from the specified windows that satisfy a waveform clipping and an amplitude outlier check. The outlier check discards records with peak ground velocities that do not lie within 6σ of the predicted value from a standard GMPE for the corresponding distance and magnitude. It is intended to discard compromised records, e.g. with grossly incorrect gain factors. In total we retain 374k 3-component quake records from 8,432 different earthquakes.

### 2.2 Impulsive Noise records ("noise")

We have used the log file data from the real-time ShakeAlert system to download waveforms around all impulsive onsets detected by the STA/LTA filter of the real-time system between June 2015 and December 2017 across the SCSN. We have removed all onset detections that occurred within 2 minutes of any earthquake in the SCEDC catalog (SCEDC, 2013), in order to avoid having real earthquake records in the noise data set. However, we note that there is still the possibility that some of these records may stem from un-cataloged local, regional and teleseismic earthquakes. The procedure results in a total number of 946k 3-component noise records.

### 2.3 Train/test splits

We split the data sets into independent training (80%) and testing data (20%) sets (Figure 1). For the quake data we split the records such that no earthquake has records in both training and testing data. This ensures full independence of the data sub-sets, and facilitates a meaningful event-by-event analysis. The noise data set is split such that training and testing subsets contain similar fractions of OnSite q-values. This is to ensure that both subsets contain similar fractions of the difficult signals, signals for which the OnSite algorithm was erroneously confident that they were local quake signals.

Most ML classifiers benefit from all classes being equally represented in the training data. Since we have a lot more noise than quake records we discard ~600k noise records, randomly selected from the noise signals with q-values of 0. Even after this removal, records with q-values of 0 make the largest share of the noise data set (Figure 1c). In total we retain 597,960 training and 149,490 validation records from the two signal classes, with 3 components each. The data sub-sets are balanced with respect to event magnitudes and hypocentral distances for the quake signals, and with respect to q-values for the noise signals.

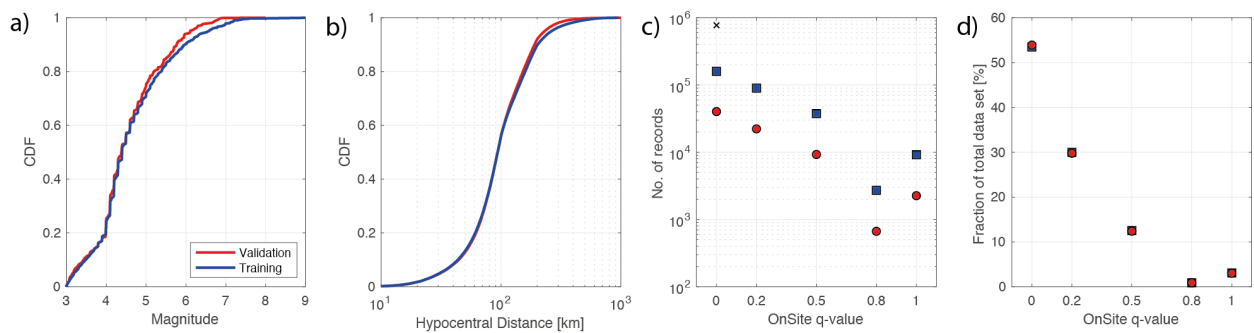

***Figure 1*** *Comparison of training and validation data sets. (a) CDFs of earthquake magnitude and (b) hypocentral distances in the quake data sets. Absolute number (c) and relative fraction (d) of records with different q-values in the noise data set. The black cross in (c) gives the*

*number of records with q-values of zero before the removal of ~600k noise records, which was done to balance the number of quake and noise records.*

## 3 Classification

We use the data to train and validate 5 different machine learning classifiers (Figure 2) for signal/noise discrimination. We evaluate the performance of the different classifiers and compare them against that of the classifier used by the OnSite algorithm (Böse et al., 2009b). While the CNN and the GAN+RF use the waveforms as model input, the other models use the pre-computed waveform features. Note that the forward computation times reported for the individual models below do not include the time to compute the features. The used features are cheap to compute, however. We do not expect that they would significantly delay real-time classifications.

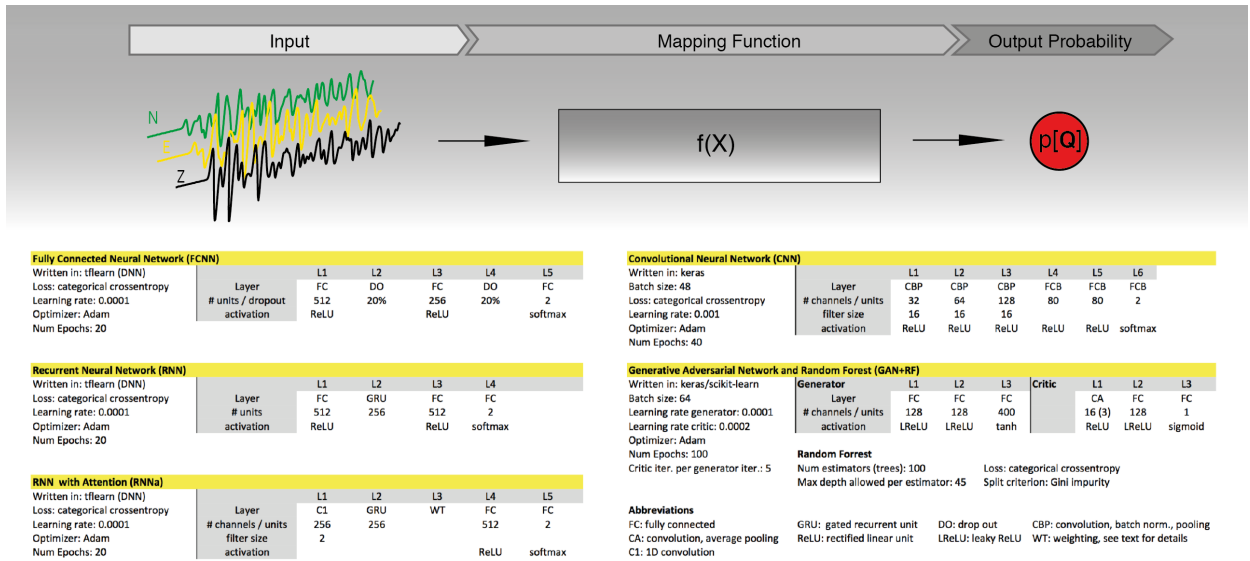

**Figure 2** *Overview and technical details of the 5 classifiers compared in this study. The classifier, f(x), acts on input features or waveforms, x, to return a probability estimate that the input signal is a local quake signal (top). Model specification for the 5 individual classifiers (bottom).*

### 3.1 Fully Connected Neural Network (FCNN)

Fully connected neural networks are a class of models that can approximate any non-linear mapping function (LeCun et al., 2015). They sequentially transform a set of input values through a large number of linear and non-linear operations into an output prediction of a target variable. In each layer of this feed-forward network, every node calculates a weighted linear transform of the outputs of all nodes in the previous layer. It then applies a nonlinear activation function (in our case ReLU for hidden layers and softmax for the final output layer) to the resulting value of the transform, which yields the output value that is passed on to the nodes in the next layer. The weights are empirically optimized to achieve maximum classification performance across the training data set. We used the Adam optimization algorithm (Kingma et al., 2014) to train the model, which is an adaptive variant of stochastic gradient descent. Dropout was applied to each of the layers with an amount set to 20% to help regularize the optimization process. The models used had three layers, composed of 512, 256, and 1 neuron (output), in order. We used the categorical cross-entropy loss function. We trained these models using an NVIDIA GTX 1080. A forward prediction took on average 3.09e-5 sec.

### 3.2 Recurrent NN (RNN)

Recurrent neural networks are a type of deep learning model that can learn temporal and contextual structure in time-series data. RNNs perform the same task for every element of the data sequence, but uses a 'repeating unit' to combine the input features of any given time step with the output from the previous time step for predictions. This adds an element of 'memory' to the net-structure of conventional, dense neural networks. RNNs can therefore draw information from how the feature values for the same site evolve as more data comes in over time. We used the gated recurrent unit (GRU) cell as our repeating unit, each with 256 nodes, and then appended a fully connected hidden layer with 512 nodes, ReLU activation and an output layer with softmax activation after the recurrent units. We used the categorical crossentropy loss function and trained the model on a NVIDIA GTX 1080 GPU. As input data we used the feature values computed on increasingly long time snippets since the signal onset, from 1-3s. A forward prediction takes on average 7.14e-5 sec.

### 3.3 Recurrent NN with Attention (RNNa)

Attention is a modification to RNNs in which, at each repeating unit, a weight vector is applied to the input vector coming from the next element in the sequence. This weight vector is unique per repeating unit, capturing the fact that different features matter more at different positions along the sequence. In the Figure 2 we describe this as an additional weighting layer, with N additional trainable parameters (the attention weights), where N is the length of the recurrent unit's output vector, for each of the T time steps (1-3s). The output of the GRU layer in the second layer (the recurrent unit) is duplicated, and this N-dimensional duplicate is element-wise multiplied with the corresponding time step's N-dimensional attention weight vector. This weighted vector and the original N-dimensional output vector are then concatenated and used as input to the next layer. The attention weights add an additional T*N trainable parameters to the network. Other than this attention modification, the parameters and training were identical to that of the plain RNN.

### 3.4 Convolutional Neural Network (CNN)

Convolutional neural networks use a set of convolutional layers in which the 3-component input waveform data is convolved with a series of parallel and sequential digital filters. After each convolution layer, the filter outputs are down-sampled and activated. The output of the last convolutional layer is fed into a fully connected network that predicts the probability that the input record is from a real earthquake. The convolutional layers act as a feature extraction system. During the training process, the digital filter coefficients and the weights from the fully connected network are jointly optimized. CNNs can achieve excellent performance, among other things, in computer vision problems, owing to their ability to detect objects in a transformation invariant manner. We use a modified version of the CNN in Ross et al., 2018b, with 3 convolutional and 2 fully connected layers. The convolutional layers consist of 32, 64 and 128 filters, respectively, with filter widths of 16. The fully connected layers consist of 80 neurons. We used ReLU activation for the hidden and softmax for the output layer. The input data were 4 seconds long waveform snippets with randomized starting times, sampled from a uniform distribution of 1.5 - 0.5 seconds before the signal onset, i.e. containing 2.5 - 3.5 seconds of actual signal. The randomized onset times ensure that the classifier does not require a very accurate onset pick. We trained it on mini-batches of 48 records, using three NVIDIA GTX 1060 GPUs, and a cross-entropy loss function. Forward prediction for a single record takes on average 1.23e-4s.

### 3.5 Generative Adversarial Network and Random Forest (GAN+RF)

GAN models (Goodfellow et al. 2014) combine a 'generator' neural network that creates synthetic data replicas, and a 'discriminator' neural network that discriminates between real and synthetic data instances. The two networks are trained to compete with each other in that the generator learns to produce increasingly realistic signals, and the discriminator learns to effectively distinguish between real and synthetic signals. Here we use the model of Li et al. [2018] that uses the discriminator network as a waveform feature extractor, and then inputs these features to a separately trained random forest classifier. We first train the GAN until the generator network can successfully generate realistic vertical P waves. In this step, only vertical earthquake P waves are input for training the GAN, and no noise data was used. Then the parameters of the first 4 layers of the discriminator network (2 convolutional layers and 2 fully connected layers) are frozen and used as the automatic feature extractor. Here the trained discriminator is assumed to have learned to recognize the key features of P waves after examining a sufficient number of training examples. We then use the discriminator to extract features from both quake and noise waveforms, and use these features to train a random forest binary classifier (100 trees with a maximum depth of 45). Note that this model uses only the vertical component of the waveforms, and uses the same randomized signal starting times and durations as the CNN. The GAN is trained on NVIDIA GTX 1050Ti GPU for 2 hours, and the Random Forest is trained for 20 min. Forward prediction for a single record takes on average 2.0e-4s.

### 3.7 The OnSite classifier

As a benchmark we use the linear signal/noise discriminator that is used in the OnSite EEW algorithm (Böse et al., 2009b). For each triggered station, OnSite computes peak absolute displacement amplitudes, and a predominant period estimate over the first few seconds of the signal (up to 4 seconds). Different combinations of these two features are associated with a probability that the record stems from a real earthquake, and are assigned a corresponding "q-value" of 0, 0.2, 0.5, 0.8 or 1. An OnSite alert is only created if the cumulative q-values across multiple stations equal at least 2. We use the q-values from the ShakeAlert log files to compare them against the false positive statistics that our non-linear classifiers produce over the same noise data set.

## 4 Classification Performance

The models are trained to distinguish the quake from the noise signals using the 598k training data records. We then apply each trained model to the independent validation data set (149k records) to evaluate the out-of-sample classification performance. For each record the models predict the probability that the target record is a real quake record. One can then define a probability threshold above which a record is considered to belong to the quake signal class. Low probability thresholds increase the likelihood of false positive cases ("FP", noise signals erroneously labeled as quake), while high thresholds increase the likelihood of false negatives ("FN", quake signals labeled as noise). The classification performance can then be analyzed in terms of how precision = TP/(TP+FP), and recall = TP/(TP+FN) vary as a function of the probability threshold. Precision/recall curves for all 5 classifiers are shown in Figure 3.

In this entire analysis we use 3-second long input signals. The FCNN model uses the features computed over the 3s long time window since the signal onset. The RNN and RNNa models use short feature time series, where the features have been computed from 1s, 2s and 3s long time windows since the signal onset. The CNN and GAN+RF models use the waveform data directly. They use on average 3s long time windows, plus 1s of pre-signal noise. The windows were uniformly randomly perturbed by +/- 0.5s around the onset such that each signal may contain between 0.5s - 1.5s of pre-pick noise data, and 2.5 - 3.5s data from after the pick.

All models show relatively high classification performance, with the deep architectures (CNN and GAN+RF) showing the lowest error rates. Thus, 3s worth of data is generally enough to discriminate between individual quake and noise records with high accuracy (e.g. 99.52% precision and 99.33% recall for the CNN). The more complex models, trained on the raw data, require larger training data sets but yield a substantial improvement over simpler models that are trained on extracted features. Among the simpler classifiers the FCNN achieves the best classification performance with recall and precision of ~99.0%. Because most records of the quake data set have not been processed by the ShakeAlert system we do not know their q-values and hence cannot evaluate the recall from OnSite. However, it is clear that all classifiers improve substantially upon the linear OnSite classifier, with complex models trained on the raw signals yielding the greatest degree of improvement.

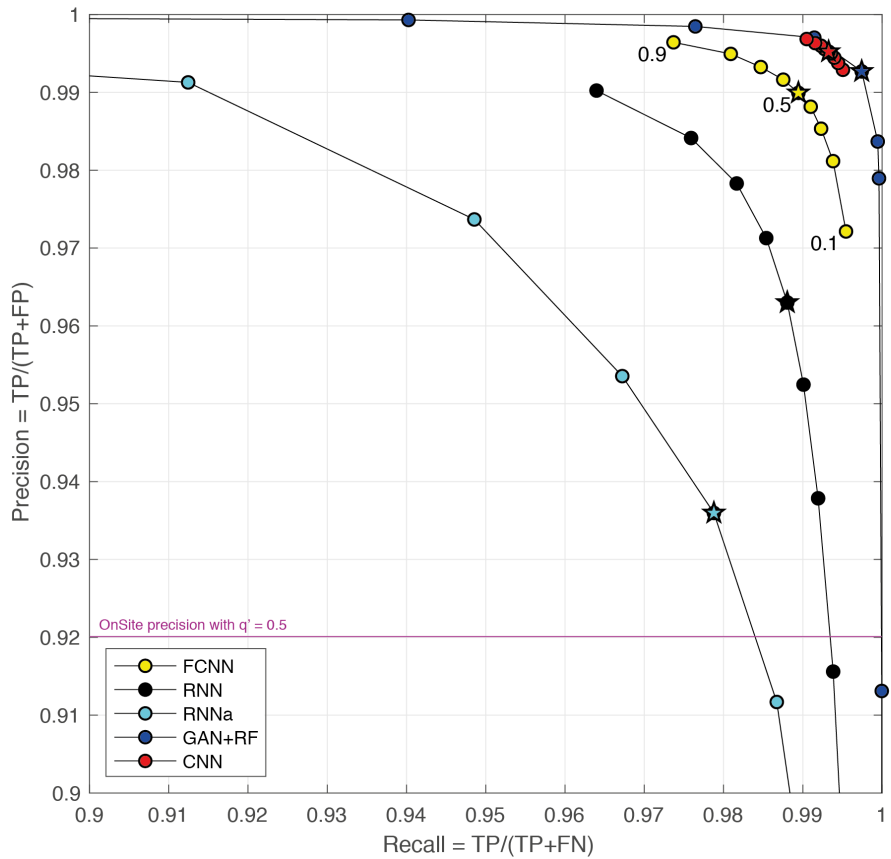

***Figure 3*** *Precision/recall curves for the 5 classifiers on the 149k validation records, using 3s long input signals. Precision and recall are computed for different probability thresholds from 0.1 and 0.9 in increments of 0.1, resulting in 9 precision and recall values for each classifier. Pentagrams show values for the probability threshold of 0.5. Models trained on raw ground motion signals have substantially higher precision and recall than those trained on extracted features. The purple horizontal line gives the precision of the OnSite classifier if records with q>=0.5 are considered quake signals.*

At first glance, the differences among the strongest models, e.g. between the FCNN and the GAN+RF are rather small and may not seem to warrant the added complexity of the deeper models. However, with the high potential trigger rates that are common for EEW systems, even small classification performance differences may have a large effect: The STA/LTA filter parameters of the ShakeAlert system (and of EEW systems in general) are set conservatively, in order to err on the side of declaring too many triggers, rather than too few. This leads to high rates of potential false triggers. The OnSite classifier assigns on average 184 non-zero q-values per day to noise signals. At such high potential false positive rates, even moderate precision improvements may substantially decrease the absolute number of false positives.

It is remarkable that, for intermediate threshold probabilities, the GAN+RF performs as well as the CNN, since it only uses the vertical component of the ground motion signals while the CNN uses all three components. A notable difference between the CNN and the other models, on the other hand, is that the precision-recall values for the CNN are much more tightly clustered. They are not a strong function of the triggering threshold. This is the case because, in a vast majority of cases, the CNN assigns probabilities close to 1 for quake records, and probabilities close to 0 for noise records (Figure 4), i.e. it is very confident in its predictions. There are very few records with intermediate probabilities that would depend on the probability threshold choice.

Since we seem to have enough data to effectively train deep classifiers, and since they perform better than the more shallow ones, we focus on the CNN and the GAN+RF in the following discussion. For reference, we also include the performance of the best among the more shallow models, i.e the FCNN.

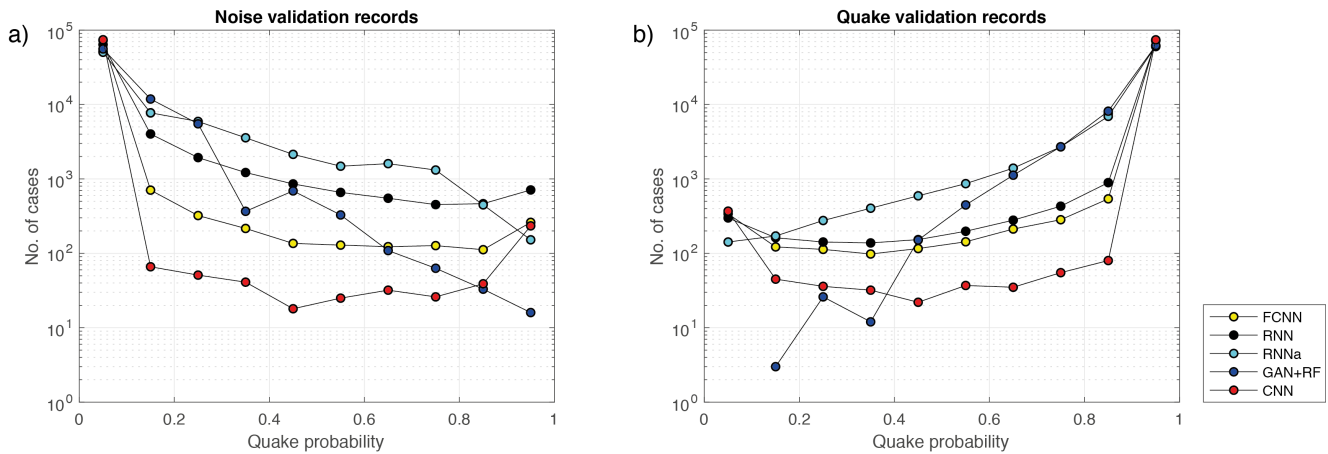

**Figure 4** *Estimated probabilities that a waveform is a real earthquake waveform from the five classifiers for noise (a) and quake (b) validation signals. While most models have thousands to tens of thousands of cases with intermediate probabilities, the CNN has only a few hundred such cases. In the rest of cases the CNN estimates probabilities ~1 for quake and ~0 for noise data. This makes it largely independent of the threshold probability choice.*

### 4.1 False Positives

If we use a probability threshold of 0.5 to classify the quake records, the CNN, GAN+RF and the FCNN classifiers produce 356, 390 and 964 FP cases from the total of 75k noise records of the validation set (0.48%, 0.52% and 1.29%, respectively). This compares to >34k non-zero q-values from the OnSite classifier on the same data set, more than 12k of which have q-values of at least 0.5. Analyzing the FP cases as a function of the OnSite q-values (Figure 5), we find that the CNN avoids >99.5% of the false triggers with high q-values, i.e. records for which the OnSite classifier was (erroneously) confident that they were real quake signals. The GAN+RF and FCNN classifiers have higher error rates for records with high q-values. The CNN classifier, on the other hand, has a higher error rate than the GAN+RF for records with q-values of zero. In summary, both of the deep classifiers would avoid a vast majority of the false positives from the OnSite classifier.

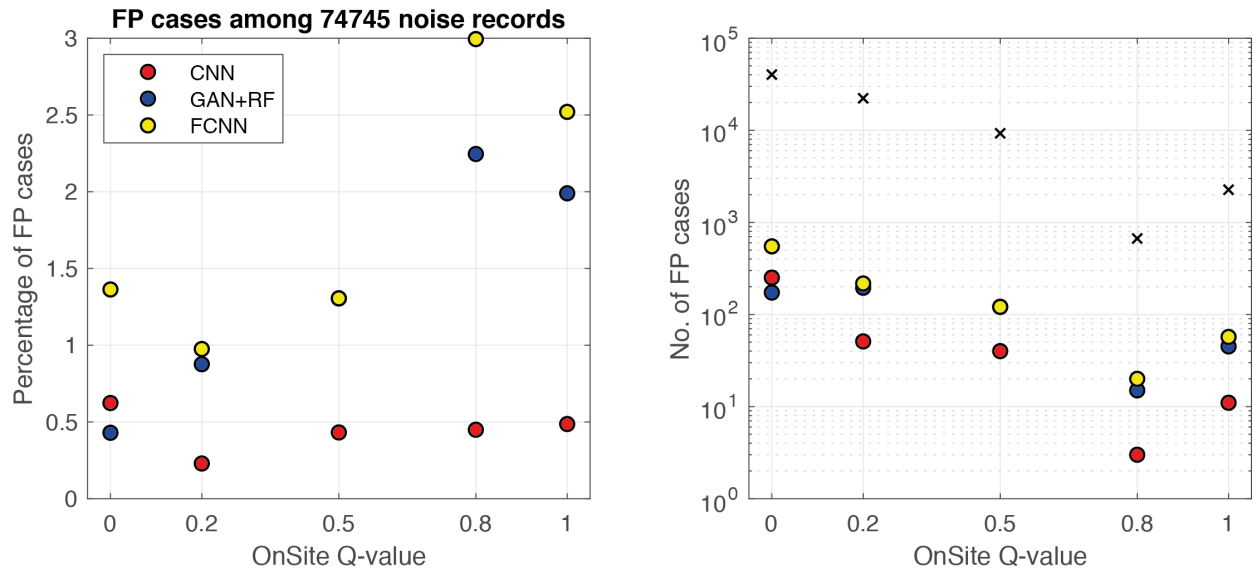

***Figure 5*** *False positive statistics for the CNN, GAN+RF and FCNN classifiers with different q-values from OnSite, with a probability threshold of 0.5. (a) Fraction of FP made by the three classifiers, relative to the number of records with a given OnSite q-value in the validation data set. (b) Absolute numbers of FP cases for each q-value. Black crosses give total number of validation records for each q-value. The CNN and GAN+RF produce only 356 and 390 FP cases, respectively, from the 75k noise records of the validation set (<0.5%). This compares to >34k records with non-zero q-values from OnSite. The CNN classifier has similar performance for records with high or low q-values, while for the GAN+RF and the FCNN classifiers the error rates increase with the q-value. The number of FPs for q-values of 0.5 for the GAN+RF and FCNN models happen to be identical.*

In the 941 days covered by the noise data set, the OnSite classifier on average made 184 false triggers per day with non-zero q-values, 65 of which have q>=0.5. If the deep classifiers can avoid 99.5% of these triggers, the typical FP rate would be at 1-2 per day on average. At such low rates it is fairly unlikely that two or more neighboring stations would happen to falsely trigger at the same time, unless they trigger on some correlated regional signal that shows up at multiple stations. An EEW system that requires a trigger at more than one station would hence almost never create false alerts because of false signal detections.

An inspection of the 356 FP signals from the CNN reveals that a majority of these signals are actually impulsive phases of teleseismic waveforms (Figure S1). The same is true for the FPs from the GAN+RF classifier. Presumably, the deep networks classify these signals as "quakes" because their particle motion is consistent with that of the local earthquake records it has been trained with. Triggering on teleseismic records is a common problem in EEW systems (e.g. Cochran et al., 2018, Hartog 2016). Phases from deep teleseisms are especially problematic as they are only subjected to crustal attenuation on their up-going path, and can have significant energy at high frequencies and thus contain impulsive phases; as a consequence they can be difficult to tell apart from local earthquakes. Furthermore, we may not have enough training data in order to effectively train a deep classifier specifically for teleseismic records. In the next section, however, we show that with a simple random forest algorithm, such signals can be reliably identified as not being local earthquakes. If we can avoid false alerts from these teleseismic signals, the FP rate of the deep networks drops to almost zero.

## 4.2 False Negatives

With a probability threshold of 0.5, the CNN, GAN+RF and FCNN models nominally produce 504, 192 and 821 FN cases, or 0.67%, 0.26% and 1.10% of the quake records of the validation set, respectively. Inspection of the falsely classified waveforms reveals that a significant fraction of these cases are mis-labeled records in the data set, i.e. there are many cases where the classifications made by the neural networks are actually correct (Figure S2). The true FN fractions are therefore even lower.

For a vast majority of earthquakes the CNN and GAN+RF missed between 0 - 10% of recordings (Figure 6), and preferentially signal with larger recording distances and correspondingly lower signal/noise ratios. This implies that these missed signal detections likely would not have led to actual missed alerts because the earthquake would still have been detected on most of the other stations. In the rare cases where the FN occur on the first few stations that should trigger, the missed triggers would have introduced some alerting delay. Thus, while in EEW missed alerts are inherently worse than false ones (Minson et al., 2018), the situation may be somewhat different on a phase-detection level: a few false detections can cause a false alert, but a few missed signal detections may merely lead to a delay, rather than to a missed alert.

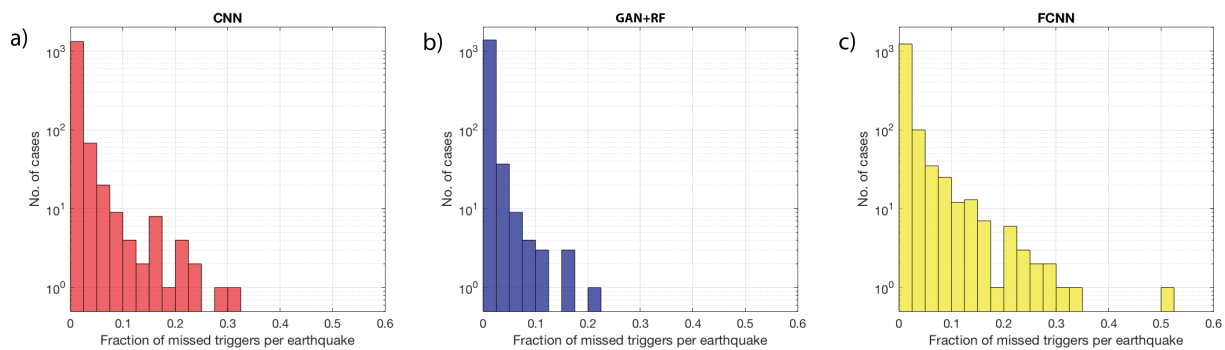

***Figure 6*** *Histograms of the fraction of FN cases (missed signal detections) for each earthquake, relative to the total number of records of these earthquakes for the CNN, GAN+RF and FCNN classifiers.*

### 4.3 Teleseismic events

In order to improve the classification performance on teleseismic signals we design and train a second classifier that could be applied to all signals that were classified as quake records by the primary classifier. The objective is to obtain a minimum number of false positives (teleseismic records classified as quakes), while having close to zero false negatives (real quake records discarded as being teleseismic records). To this end we compile a separate data set that consists of teleseismic signals that were impulsive enough to trigger the STA/LTA filter of the operational ShakeAlert system.

We used the Seismic Transfer Protocol (STP) to download all teleseismic records from SCSN stations from January 2005 to December 2016. We then run the STA/LTA trigger from OnSite on these waveforms. STP returns windowed waveforms irrespective of whether or not there is a discernible teleseismic signal in the data, and often there is not. Furthermore, for many of the stations, there is a high rate of false STA/LTA triggers. As a consequence, a majority of these picks are actually random noise triggers, not teleseismic ones. To single out the teleseismic waveforms we require for each event that at least 15 phases line up in terms of their phase move-out. This leads to 7,544 3-component records of which we are confident that they are teleseismic waveforms. Lowering the minimum number of phases to below 15 would increase the number of records, but it would come at the cost of contaminating the data set with non-teleseismic signals. We apply the same processing as we did to the other two signal classes and extract 25% of the data set for model validation. Our splits are designed such that no earthquake has records in both training and testing data, as was done for the quake data set.

This data set is not large enough to effectively train a deep network classifier (6,035 training and 1,509 3-component validation records). Instead, we train a simple random forest classifier, making use of the pre-computed features listed in Table 1. Because random forests can still perform relatively well even with large class imbalance, we were able to use vastly more quake training records (for which we have a much bigger set) than teleseismic training records. We use all 231,482 records with hypocentral distances <150km from the quake training set. The large number of quake records helps the model to learn the full breadth of signal forms that quake records can come in, and leads to vastly better classification results than when the quake set is trimmed to the size of the teleseismic data set.

With a maximum tree depth of 40 and 100 trees we achieve a classification accuracy for the quake data of 99.95% (i.e. almost no quake records discarded), and 95.36% for the teleseismic records. Thus, although this removes most of the teleseismic triggers, occasional false triggers from impulsive teleseismic records can still occur and pose a problem to EEW algorithms. EEW developers may have to resort to multi-station logic (e.g. requiring an upper limit on apparent phase velocities) to ensure that they never trigger on teleseismic records. Alternatively, larger data sets of impulsive teleseismic records may in the future enable the training of deeper classifiers with potentially higher classification performance.

**5 Discussion**

Our analysis demonstrates that machine learning algorithms allow us to build highly accurate classifiers that can strongly benefit EEW algorithms. The classifiers are able to reliably discriminate EEW-relevant local earthquake signals from a vast range of non-relevant nuisance signals. The volume of currently available seismic data is sufficient to train complex, deep classifiers that reach accuracies of ~99.5% on independent validation data, and that can be run in real-time. Such machine learning based classifiers have the potential to significantly reduce false and missed alert occurrences in the next generation EEW systems. Furthermore, better signal/noise discrimination can also speed up the alerting: with the reliability of the discussed classifiers, EEW systems may be able to send out reliable alerts based on the signals of only one or two stations. This would save the time it takes for the signals to travel to more distant stations, and may reduce the size of the EEW blind zone. At the same time, despite being faster, the alerts may be more reliable than the current 4- or more station alerts.

Another important advantage from using deep learning classifiers for EEW signal detection is that they are much better at detecting signals during periods of high noise amplitudes, e.g. during an aftershock sequence. Standard detection algorithms that are based just on signal amplitudes (in particular the commonly used STA/LTA detectors) lose most of their detection sensitivity once a high amplitude signal raises the baseline level. The deep learning classifiers, on the other hand, can detect seismic signals even in the presence of high noise levels, e.g. if they fall into the waveform coda of a previous event (Ross et al., 2018b).Our comparison of the five classifiers demonstrates that the more complex models trained on raw ground motion signals (CNN and GAN+RF) significantly outperform the simpler classifiers that were trained on hand-engineered waveform features. Ross et al., 2018b have previously shown that the same type of CNN classifiers excel at the task of discriminating seismic P- and S-phases recorded at short distances (<100km), and non-impulsive ambient background noise signals that had been collected from a few seconds before P-phase arrivals. Our analysis here demonstrates that we can reach similar performance against more earthquake-like impulsive noise signals, and across a much wider distance (out to 1,000km) and magnitude (M3.0-9.0) ranges. The classification performance achieved in this work also demonstrates that a single model should be sufficient for signal/noise discrimination for all stations in an EEW network. The individual recording sites do not seem to differ enough in terms of their local site and noise characteristics such that a station-wise model training would be necessary.

With appropriate GPU hardware all models discussed in this study are fast enough for simultaneous processing of a large number of channels in real-time. An interesting future extension of these models would be to train more shallow classifiers that can be run on cheaper instruments, e.g. on phone accelerometers (e.g. Kong et al., 2016) and MEMS accelerometers (e.g. Cochran et al., 2009), or to use consumer grade GPUs to run deeper models directly on site. Machine learning classifiers could also be incorporated into more standard single-site EEW systems (e.g. Hsu et al., 2016, GRL), which are simpler to run than network-based EEW systems, and which can potentially provide timely alerts for sites that are in the blind zone of the network-based systems. Since such systems cannot resort to multi-stage logic in their event detection, reliable signal/noise discrimination is all the more important.

It is important to note that, while machine learning based classifiers can avoid the vast majority of false triggers on an individual station level, false EEW alerts may still occur. This can happen for a variety of reasons, such as errors in phase associations or overestimations of small magnitude events. However, the signal/noise discrimination achieved in this work is undoubtedly an important step towards building more robust EEW systems. Furthermore, different deep learning algorithms may also provide more reliable solutions for these other aspects of real-time seismic monitoring, e.g. for seismic phase association (Ross et al., 2018c).

The main caveat of the presented deep models is their susceptibility to teleseismic signals. The limited size of our training dataset does not yet allow efficient training of deep models with teleseismic signals as an individual class. Larger teleseismic data compilations may in the future allow harnessing the power of deep classifiers also for this model class. Another caveat is that, if the models are applied to data from new stations with special noise characteristics that are not well represented in the noise training data, the classification performance may somewhat degrade. For such stations, a model may have to be retrained after adding noise from that particular station to the training data. For stations that have more typical noise signal characteristics (and hence are well represented in the training data set) the trained model should be directly applicable. Our results suggest that the deep models perform similarly in detecting Japanese and Californian earthquakes: the CNN, for instance, has a false negative rate of 1.45% for the quake data from southern California, and 0.34% for the Japanese quake data. The difference may come from the larger Japanese share of the quake data set (70%), or because the Japanese data may be cleaner and generally of better quality. Since we do not have impulsive noise recordings from Japan we cannot estimate the regional differences in the false positive rates. A quantitative assessment of the portability of these classifiers to new stations and/or new target regions will be the focus of dedicated future studies.

# 6 Conclusions

Our results demonstrate that machine learning classifiers hold vast promise for making EEW systems faster and more reliable. The currently available amount of seismic waveform data is sufficient to train deep learning classifiers that need only a few seconds of signal to discriminate between earthquake and nuisance signals in real-time, and with high reliability (accuracy of 99.5%). With the classification performance that the deep learning classifiers achieve, it may be possible to send out EEW alerts based on information from only one or two stations, even in networks with high-noise stations. This would significantly speed up the alerting, and may allow to provide timely warning to near-epicentral sites where strong ground motion sets in fast, and where the alerts are most strongly needed.

**Acknowledgments**

This research was supported by a Gordon and Betty Moore Foundation grant to Caltech and by the Swiss National Science Foundation. The Japanese waveform data can be downloaded from http://www.kik.bosai.go.jp/ (last accessed October 2017). For southern California we have used waveforms and parametric and waveform data from the Caltech/USGS Southern California Seismic Network (doi:10.7914/SN/CI) stored at the Southern California Earthquake Data Center (doi:10.7909/C3WD3xH1). The waveform and feature data set is available as a single hdf5 file at scsn.org/xx/. The algorithms were written with Python packages TensorFlow (https://tensorflow.org/), TFLearn (http://tflearn.org/), Keras (https://keras.io/) and Scikit-learn (http://scikit-learn.org/).

## References

Böse, M., Wenzel, F. and Erdik, M., 2008. PreSEIS: A neural network-based approach to earthquake early warning for finite faults. *Bull. Seism. Soc. Am.*, *98*(1), pp.366-382.

Böse, M., Hauksson, E., Solanki, K., Kanamori, H., and Heaton, T.H., 2009a. Real-Time Testing of the On-site Warning Algorithm in Southern California and Its Performance During the July 29 2008 Mw5.4 Chino Hills Earthquake, *Geophys. Res. Lett.*, Vol. 36, L00B03, doi:10.1029/2008GL036366.

Böse, M., Hauksson, E., Solanki, K., Kanamori, H., Wu, Y.-M., and Heaton, T.H., 2009b. A New Trigger Criterion for Improved Real-time Performance of On-site Earthquake Early Warning in Southern California, *Bull. Seism. Soc. Am.*, 99, 2-A, pp. 897-905, doi: 10.1785/0120080034.

Chen, Y., 2017. Automatic microseismic event picking via unsupervised machine learning. *Geophysical Journal International*, *212*(1), pp.88-102.

Cochran, E.S., Lawrence, J.F., Christensen, C. and Jakka, R.S., 2009. The quake-catcher network: Citizen science expanding seismic horizons. *Seismol. Res. Lett.*, *80*(1), pp.26-30.

Cochran, E.S., Kohler, M.D., Given, D.D., Guiwits, S., Andrews, J., Meier, M.A., Ahmad, M., Henson, I., Hartog, R. and Smith, D., 2017. Earthquake early warning ShakeAlert system: Testing and certification platform. *Seismol. Res. Lett.*, *89*(1), pp.108-117.

Deng, J., W. Dong, R. Socher, L.-J. Li, K. Li and L. Fei-Fei, 2009. ImageNet: A Large-Scale Hierarchical Image Database. *IEEE Computer Vision and Pattern Recognition (CVPR)*.

Donahue, J., Hendricks, L.A., Guadarrama, S., Rohrbach, M., Venugopalan, S., Saenko, K., Darrell, T., 2015. Long-term recurrent convolutional networks for visual recognition and description. *The IEEE Conference on Computer Vision and Pattern Recognition (CVPR)*, pp. 2625-2634

Given, D.D., Cochran, E.S., Heaton, T., Hauksson, E., Allen, R., Hellweg, P., Vidale, J. and Bodin, P., 2014. Technical implementation plan for the ShakeAlert production system: An earthquake early warning system for the west coast of the United States (No. 2014-1097). US Geological Survey.

Goodfellow, I., Pouget-Abadie, J., Mirza, M., Xu, B., Warde-Farley, D., Ozair, S., Courville, A. and Bengio, Y., 2014. Generative adversarial nets. *Advances in neural information processing systems* (pp. 2672-2680).

Hammer, C., Beyreuther, M. and Ohrnberger, M., 2012. A seismic‐event spotting system for volcano fast‐response systems. *Bull. Seism. Soc. Am.*, *102*(3), pp.948-960.

Hartog, R., J., Kress, V.C., Malone, S.D., Bodin, P., Vidale, J.E. and Crowell, B.W., 2016. Earthquake early warning: ShakeAlert in the Pacific Northwest. *Bull. Seism. Soc. Am.*, *106*(4), pp.1875-1886.

Hsu, T.Y., Wang, H.H., Lin, P.Y., Lin, C.M., Kuo, C.H. and Wen, K.L., 2016. Performance of the NCREE's on-site warning system during the 5 February 2016 Mw 6.53 Meinong earthquake. *Geophys. Res. Lett.*, *43*(17), pp.8954-8959.

Kingma, Diederik P and Ba, Jimmy Lei, 2014. Adam: A method for stochastic optimization. arXiv preprint arXiv:1412.6980.

Kohler, M.D., Cochran, E.S., Given, D., Guiwits, S., Neuhauser, D., Henson, I., Hartog, R., Bodin, P., Kress, V., Thompson, S. and Felizardo, C., 2017. Earthquake early warning ShakeAlert system: West coast wide Production Prototype. *Seismol. Res. Lett.*, *89*(1), pp.99-107.

Kong, Q., Allen, R.M., Schreier, L. and Kwon, Y.W., 2016. MyShake: A smartphone seismic network for earthquake early warning and beyond. *Sci. Adv.*, *2*(2), p.e1501055.

LeCun, Y., Bengio, Y. & Hinton, G., 2015. Deep learning. Nature 521, 436–444.

Levine, S., Finn, C., Darrell, T., and Abbeel, P, 2016. End-to-end training of deep visuomotor policies. *JMLR 17*.

Li, Z., Meier, M.A., Hauksson, E., Zhan, Z. and Andrews, J., 2018. Machine Learning Seismic Wave Discrimination: Application to Earthquake Early Warning. *Geophys. Res. Lett.*.

Maggi, A., Ferrazzini, V., Hibert, C., Beauducel, F., Boissier, P. and Amemoutou, A., 2017. Implementation of a multistation approach for automated event classification at Piton de la Fournaise volcano. *Seismol. Res. Lett.*, *88*(3), pp.878-891.

Meier, M.A., 2017. How "good" are real-time ground motion predictions from earthquake early warning systems?. *Journal of Geophysical Research: Solid Earth*, *122*(7), pp.5561-5577.

Minson, S.E., Meier, M.A., Baltay, A.S., Hanks, T.C. and Cochran, E.S., 2018. The limits of earthquake early warning: Timeliness of ground motion estimates. *Sci. Adv.*, *4*(3), p.eaaq0504.

Mousavi, S.M., Horton, S.P., Langston, C.A. and Samei, B., 2016. Seismic features and automatic discrimination of deep and shallow induced-microearthquakes using neural network and logistic regression. *Geophysical Journal International*, *207*(1), pp.29-46.

Panakkat, A. and Adeli, H., 2009. Recurrent neural network for approximate earthquake time and location prediction using multiple seismicity indicators. *Computer‐Aided Civil and Infrastructure Engineering*, *24*(4), pp.280-292.

Perol, T., Gharbi, M., Denolle, M., 2018. Convolutional neural network for earthquake detection and location. *Sci. Adv.* 4, e1700578.

Peters, M.E., Neumann, M., Iyyer, M., Gardner, M., Clark, C., Lee, K., and Zettlemoyer, L. Deep contextualized word representations, 2018. *In Proceedings of NAACL*.

Rong, K., Yoon, C.E., Bergen, K.J., Elezabi, H., Bailis, P., Levis, P. and Beroza, G.C., 2018. Locality-Sensitive Hashing for Earthquake Detection: A Case Study Scaling Data-Driven Science. *arXiv preprint arXiv:1803.09835*.

Rouet-Leduc, B., Hulbert, C., Lubbers, N., Barros, K., Humphreys, C. J., & Johnson, P. A., 2017. Machine learning predicts laboratory earthquakes. *Geophys. Res. Lett.*, *44*, 9276–9282. https://doi.org/10.1002/2017GL074677

Ross, Z.E., Meier, M.-A. and Hauksson, E., 2018a. P-wave arrival picking and first-motion polarity determination with deep learning. *Journal of Geophysical Research: Solid Earth*.

Ross, Z.E., Meier, M.-A., Hauksson, E., and Heaton, T. H., 2018b. Generalized Seismic Phase Detection with Deep Learning. *Bull. Seism. Soc. Am.,* doi: https://doi.org/10.1785/0120180080

Ross, Z.E., Yue, Y., Meier, M.-A., Hauksson, E. and Heaton, T.H., 2018c. PhaseLink: A Deep Learning Approach to Seismic Phase Association. *arXiv preprint arXiv:1809.02880*.

Sermanet, P., Eigen, D., Zhang, X., Mathieu, M., Fergus, R. and LeCun, Y., 2013. Overfeat: Integrated recognition, localization and detection using convolutional networks. *arXiv preprint arXiv:1312.6229*.

SCEDC (2013): Southern California Earthquake Center. Caltech.Dataset. doi:10.7909/C3WD3xH1

Trugman, D.T. and Shearer, P.M., 2018. Strong Correlation between Stress Drop and Peak Ground Acceleration for Recent M 1–4 Earthquakes in the San Francisco Bay Area. *Bull. Seism. Soc. Am.*, *108*(2), pp.929-945.

Valentine, A.P. and Trampert, J., 2012. Data space reduction, quality assessment and searching of seismograms: autoencoder networks for waveform data. *Geophysical Journal International*, *189*(2), pp.1183-1202.

Wang, J., Teng, T.-L., 1995. Artificial neural network-based seismic detector. *Bull. Seismol. Soc.Am*. 85, 308–319.

Xu, Y., Wang, J.P., Wu, Y.M. and Kuo-Chen, H., 2017. Reliability assessment on earthquake early warning: A case study from Taiwan. *Soil Dynamics and Earthquake Engineering*, *92*, pp.397-407. http://dx.doi.org/10.1016/j.soildyn.2016.10.015

C. E. Yoon, O. O'Reilly, K. J. Bergen, G. C. Beroza, 2015. Earthquake detection through computationally efficient similarity search. *Sci. Adv.* 1, e1501057.

Zhu, W. and Beroza, G.C., 2018. PhaseNet: A Deep-Neural-Network-Based Seismic Arrival Time Picking Method. *arXiv preprint arXiv:1803.03211*.

Zhu, Y., Mottaghi, R., Kolve, E., Lim, J. J., Gupta, A., Li, F.-F., and Farhadi, A., 2017. Target-driven visual navigation in indoor scenes using deep reinforcement learning. *In IEEE International Conference on Robotics and Automation (ICRA)*.